\documentclass[10 pt]{amsart}

\usepackage{amsmath,amssymb,amscd,latexsym,euscript,mathrsfs}

 \usepackage{graphicx}

\def\leq{\leqslant}
\def\geq{\geqslant}

\begin{document}

\title{Optimum Depth of the Bounded Pipeline}

~\\

\noindent
{\it A. A. Husainov}
\footnote{Husainov Ahmet Aksanovich - doctor of physical and 
mathematical sciences, professor; 
husainov51@yandex.ru}

~\\

\noindent
\large{\bf OPTIMUM DEPTH OF THE BOUNDED PIPELINE}
\\
~\\
\noindent
{Komsomolsk-on-Amur State University, 27, Lenina prosp., 
Komsomolsk-on-Amur, 681013, Russian Federation}

~\\

The paper is devoted to studying the performance of a computational pipeline, the number
of simultaneously executing stages of which at each time is bounded from above by a fixed number.
A look at the restriction as a structural hazard makes it possible 
to construct an analytical model for calculating the processing time 
of a given input data amount.
Using  this model, led to a formula for calculating the optimal depth of a bounded pipeline for a given volume of input data.
The formula shows that the optimal 
depth can get large changes for small changes in the amount of data. 
 To eliminate this disadvantage and to obtain a more convenient formula for optimal depth, 
 a pipeline with a single random hazard is constructed, the mathematical expectation 
 of a random value of the processing time of which approximates the analytical model of the bounded pipeline. 
In addition, a pipeline with two hazards has been built, the analytical model of which allowed 
 obtaining formulas for calculating the optimal depth of a bounded pipeline with restart 
 for a given amount of data. 
 To check whether the proposed analytical models are consistent with the experiments to calculate the processing time, two methods of computer simulation of bounded pipelines are used, the first of which is constructed as a multi-threaded application, and the second is based on the theory of free partially commutative monoids.

~\\



~\\

\section{Introduction}

The stages of the computational pipeline operating at a given time are called 
\emph{active}.
A computational pipeline is called \emph{bounded} if, in each 
moment of time, the number of its active stages is no greater 
than a some fixed integer $q\geq 1$.
The problem of evaluating the performance of 
 a bounded pipeline can be 
encountered when using multi-threaded pipelines, 
in which threads running on a multi-core processor with common 
memory perform the role of stages. At each time, the number of its active 
stages is bounded by the number of processor cores.
A similar problem can arise when there are not enough other resources, 
for example, when each active stage uses one of the channels of a multi-channel
memory controller. In this case, the number of active stages of the pipeline
bounded by the number of channels of the controller.

A bounded pipeline can be viewed as a sequence of stages, each of which 
has a register (latch) for writing the result of the stage operation, and a set 
of $q$ functional devices serving the stages. Hence it follows that the acceleration 
of a bounded pipeline can be estimated using the methods of Amdahl \cite{amd1967}.
The idea of considering segments separately 
from these devices is not new and is described in detail, for example, 
in \cite{mor2012}-\cite{mor2017}. 
On the basis of experiments, in the form of a conjecture, 
an analytical model was proposed in \cite{X2016} for calculating 
the time of processing data of a given amount using a bounded pipeline.
It follows from Proposition 1, proved bellow, that this connjecture is not true in general, 
but in the case of a uniform pipeline it gives a good approximation for the processing 
time of the given amount. 

The \emph{depth} of the pipeline is the number of its stages.
The depth is \emph{optimal} if the processing time of input data 
is minimal. 

The purpose of this paper is to find formulas for calculating the depth
of a uniform bounded pipeline 
for a given amount of input data.

Historically, various models have been used to calculate the optimum depth.
Using the model for \emph{throughput} (see \cite{fly1999}), Dubey and Flynn \cite{dub1990}
obtained the formula for the optimal depth for the pipeline with restart. 
Emma and Davidson \cite{emm1987} used to calculate the \emph{inverse banwidth}
and showed that, in general, the optimal depth can be characterized by 
$p_{opt}= \sqrt{\gamma\alpha}$ where $\gamma=\frac{t_p}{t_o}$
is the ratio of overall circuit delay $t_p$ to latching overhead $t_o$,
and $\alpha$ is a function of trace statistics that accounts 
for the delays induced by data dependencies and branches.
Interesting model and a general formula for optimal 
depth were obtained by Hartstein and Puzak \cite{har2002}. In \cite{X20182}, a refinement 
of the Dubey and Flynn formula, taking into account the data amount, was obtained.
 We find similar formulas for bounded pipelines.

In this paper,
we note that limiting the number of active stages leads  
to a structural hazard. This view allows us to construct an analytical model
for calculating  
the processing time of a given amount of data (Proposition 1). 
The analytical model leads to one of the main results of the paper, to the formula for calculating the optimal depth (Theorem 1).
It shows that the optimal number of stages can be greater than the 
number of active stages.
We find that the depth obtained may depend too much on the amount of data.
For those cases where the amount of data can vary,
 we propose to consider a bounded pipeline as a \emph{simplified} pipeline with
 a single random hazard. 
The mathematical expectation of a random time value 
of processing time of the simplified pipeline approximates the analytical model of the
 bounded pipeline. 
The accuracy of the approximation describes Proposition 2. 
A formula is proposed for the optimum depth of a bounded 
pipeline (Corollary 1). 
To find the optimal depth of the bounded pipeline with restarting, 
a pipeline with two random hazards is considered (Theorem 2). 
The final part of the paper is devoted to two methods of computer simulating 
the bounded pipelines. The first method is based on 
a multi-threaded application in which threads play the role of stages. The second
 is based on the algorithm for reducing the trace consisting of operations 
 to the Foata normal form 
\cite{die1990}.

\section{Bounded pipeline}

The pipeline stage has a storage device and is connected with two registers (\emph{latches}), 
one of which is called the \emph{input}, and the other is the \emph{output} register. 
The stage consists of three operations: reading data from the input register, 
a logical stage operation, and writing data to the output register. 
 A stage can have a local memory 
to store an internal state. The \emph{functional device} is intended for servicing the stage.
It performs all three steps of the stage. The time of a one-time run of a stage is called 
the \emph{stage delay}.
The \emph{bounded pipeline} consists of 
$p$ stages and $p+1$ latches, 
and a set of $q\leq p$ functional devices. The sequence of segments and latches is connected 
as follows:
$$
latch_0\to stage_1\to latch_1\to \ldots \to stage_p \to latch_p.
$$
Arrows indicate the direction of data transfer. 
The input elements are entered into the pipeline by $latch_0$. 
At any time, the input element can be processed by no more than one functional device. 
Each stage has a set of functional devices capable of executing its stage operations. 
A stage is called {\it active} at a given time, if it is at that time served by one 
of the functional devices. Several stages can be active, but not more than $q$. 

A bounded pipeline is called \emph{uniform} if all its stages have the same delays.
Throughout the paper, we consider uniform bounded pipelines. 
The stage delay is called the \emph{pipeline cycle} 
$h= \frac{t_p}{p}+t_o$ where $t_p$ is the \emph{logical delay} of a pipeline equal to the time of sequential execution of operations of all stages except of input/output operations for latches, and $t_o$ is the time of input/output operations for the stage. 

Below everywhere, $k {\rm~mod~}{q}$ denotes the remainder of dividing a nonnegative integer $k$
 by a natural number $q\geq 2$. A bounded pipeline can be implemented 
in digital signal processing processors \cite[Section 4.2]{kog1981}. 
For example, this can be done for the case when the number of stages $p$
 is a multiple of the number of functional devices $q$, and any stage 
with the number $1\leq k\leq p$ has a functional device with the number
$ 1 + k {\rm~mod~}{q}$ that is capable of executing the operation of the stage. 
Then there will be a strong inequality $q<p$, but the number of latch registers 
 will not change.

If $p=q$, then there are no hazards, and the processing 
time of $n$ elements equals $(p + n-1) h$, where $n$ is the number of input elements of the pipeline.

Let $p$ be the depth of a pipeline and let $q$ be the number of 
functional devices  
such that $1\leq q\leq p$. The input of the pipeline receives $n$ 
elements of input data. The first stage performs the first operation on each 
of these elements and transfers the results for the second stage. 
The second stage receives these results, performs its operation and transfers 
the results for the third, etc. If we try to start more than $q$ parallel stages, 
a structural conflict occurs.  As a result, each inactive stage will wait 
for the release of one of the functional devices, and the operating time of this 
stage will increase by $(p-q) h$.

Under the \emph{reservation table} \cite{kog1981} of a pipeline, we mean a matrix whose element 
$a_{i j} = k$ if and only if the
 $i$-th stage processes the $k$-th input element at time $j$. 
Table 1 shows a reservation table for 
a bounded pipeline consisting of $4$ stages and $3$ functional devices 
processing $8$ input elements.

\begin{table}
{\it Table 1.} {\bf Reservation table of the pipeline}

\begin{tabular}[t]{|c|c|c|c|c|c|c|c|c|c|c|c|c|c|}
\hline
 & \tiny{01} & \tiny{02} & \tiny{03} & \tiny{04} & \tiny{05} & 
 \tiny{06} & \tiny{07} & \tiny{08} & \tiny{09} & \tiny{10} &
 \tiny{11} & \tiny{12} & \tiny{13} \\
 \hline
\small{1} & \small{1} & \small{2} & \small{3} &  & \small{4} & \small{5} 
& \small{6} & & \small{7} & \small{8} & & & \\
\hline
\small{2} & & \small{1} & \small{2} & \small{3} &  & \small{4} & \small{5} 
& \small{6} & & \small{7} & \small{8} & &\\
\hline
\small{3} & & & \small{1} & \small{2} & \small{3} &  & \small{4} & \small{5} 
& \small{6} & & \small{7} & \small{8} & \\
\hline
\small{4} & & & & \small{1} & \small{2} & \small{3} &  & \small{4} & \small{5} 
& \small{6} & & \small{7} & \small{8} \\
\hline
\end{tabular}
\end{table}

Hence we arrive at the following formula for the processing time of $n$ elements using $q$ processors 
for a uniform pipeline of $p$ stages.
\begin{equation*} 
T_q (p, n) = \left(p + n-1 + (p-q) ^ + \left[\frac{n -1}{q}\right]\right) h,
\end{equation*}
where $[x]$ is denoted the integer part of $x$ and
$x^+$ is a function of $x$ such that $x^+ = x$ if $x\geq 0$, and $x^+ = 0$ otherwise.  
Substituting in the obtained formula the delay of the stage $h = \frac{t_p}{p} + t_o$, 
we arrive at the following assertion:

\newtheorem{prop1}{Proposition}
\begin{prop1}
The processing time of $n$ elements using a pipeline of depth $p$, the number
 of active stages of which is bounded at each time by the number $q\geq 1$, 
is equal to $T_q (p, n) = \left(p + n-1 + (p-q)^+ \left[\frac{(n -1)}{q}\right]\right)
 (\frac{t_p}{p} + t_o)$.
\end{prop1}

We note that if we use the Amdahl formula \cite{amd1967} for the calculation time $n p$
 of operations, then the approximation obtained will be good, but not exact. 
The correct formula can be obtained with the help of the 
Generalized Amdahl Law 
$T_q (n) = (\frac{g_1}{1} + \frac{g_2}{2} + \ldots + \frac{g_q}{q}) T_1 (n)$
 from the monograph \cite{she2005}, where $g_i$ represents the fraction of time
when there are
 $i$ concurrently processing stage operations. 
In the example shown in Table 1, we have $g_1 = \frac{2}{32}$, $g_2 = \frac{6}{32}$, 
$g_3 = \frac{24}{32}$.

\section{The optimal depth of a bounded pipeline for a given amount of input data}

Given the number of functional devices $q$ and the data amount $n$, 
the optimal depth $p_{opt} (n, q)$ of the bounded pipeline
is the number of stages $p$, at which the time $T_q (p, n)$ is minimal. 
The graph of the curve $y = T_q (x, n)$ consists
of two parts of hyperbolas (Fig. 1).
It illustrates the dependence of the processing time
at the number of stages $x$ for 
 $n = 20$ of input elements, $t_o = 0.3$, $t_p = 1$, $q = 5$. 
The graph is marked by a thick line.

\begin{center}
\includegraphics[height=50mm]{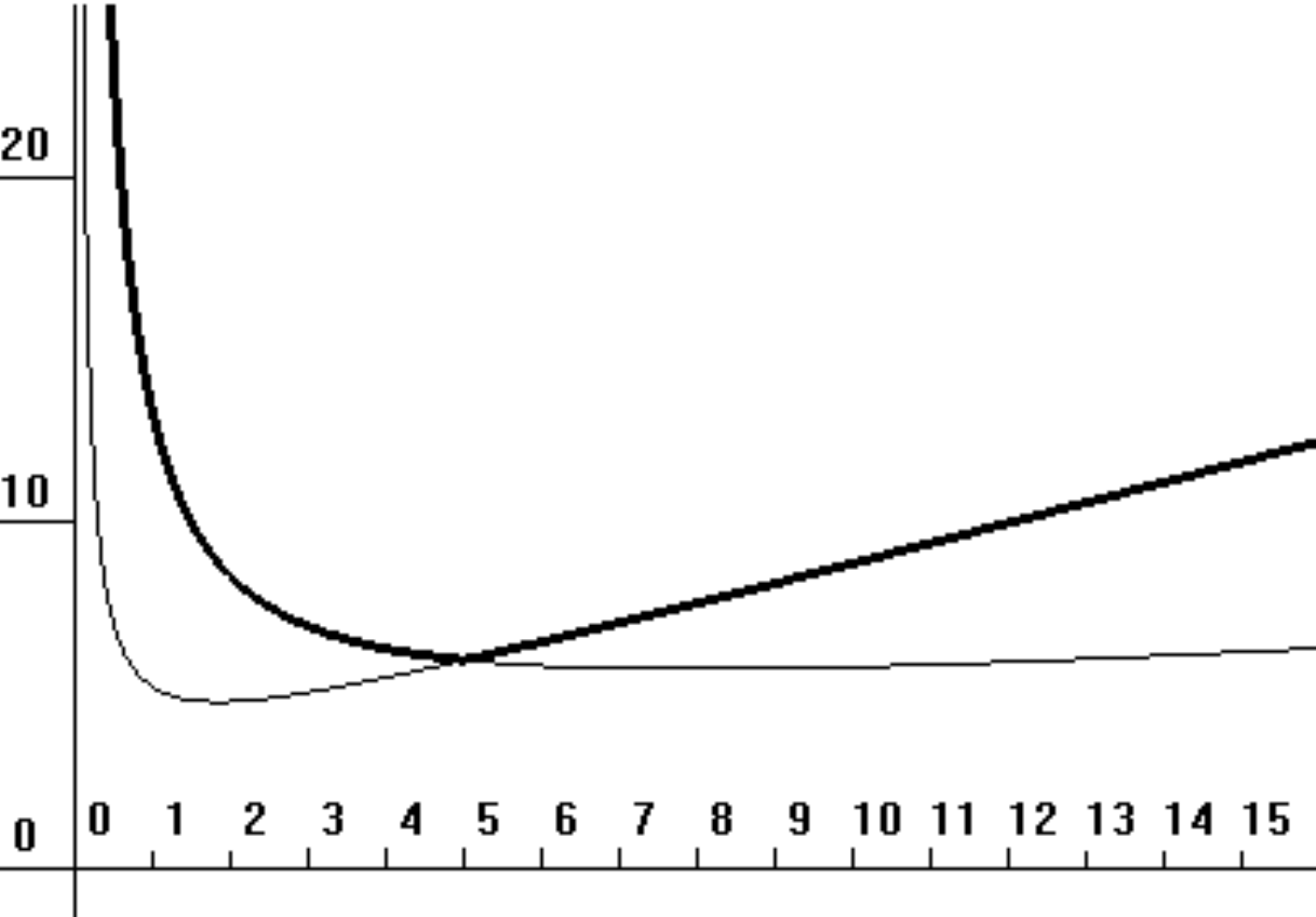}

{\it Fig. 1.} {Graph of the dependence of the processing time at the number of stages}
\end{center}

To study the graph of the function $y = T_q (x, n)$ on the number of stages $x$, 
for constants $q$ and $n$, we remark that for each 
function $f (x) = A x + B + \frac{C}{x}$, all of whose 
coefficients $A, B< C$ are nonnegative, there are the following cases: If $C \geq 0$, 
then this function has the vertical asymptote $x = 0$ and the asymptote $y = Ax + B$. 
If $A>0$, then the function decreases and reaches a minimum at 
$x = \sqrt{\frac{C}{A}}$. If $A = 0$, then the function decreases for $x\to \infty$, 
in this case its limit value is equal to $B$.
For $C = 0$, we have $f (x) = Ax + B$.

Recall that $(n-1){\rm~mod~}{q}$ 
denotes the remainder of dividing the number $(n-1)$ by $q$.
Consider $T_q (x, n)$ as a function of $x> 0$, for fixed $q$ and $n$. 
The graph of this function consists of points belonging to two hyperbolas. 
The first hyperbola consists of points $(x, f_q (x))$, where
$$
f_q (x) = \left(x + n-1 + (x-q) \left[\frac{n-1}{q}\right]\right) (t_o + \frac{t_p}{x}),
$$ 
for all $x> 0$,
and the second from the points $(x, f (x))$, where 
$f (x) = (x + n-1) (t_o + \frac{t_p}{x})$, for all $x> 0$.
It is easy to see that $f_q (x) = A_qx + B_q + \frac{C_q}{x}$, where 
$A_q = t_o \left(1 + \left[\frac{n-1}{q}\right]\right)> 0, 
B_q = t_o ((n-1) {\rm~mod~}{q}) + t_p \left(1 + \left[\frac{n-1}{q}\right]\right)> 0, 
C_q = t_p ((n-1){\rm~mod~}{q}) \geq 0$. 
Similarly, $f (x) = Ax + B + \frac{C}{x}$, 
where $A = t_o> 0$, $B = (n-1) t_o + t_p> 0$, $C = (n-1) t_p> 0$. 
These hyperbolas have a common vertical asymptote $x = 0$. 
The first hyperbola also has the asymptote $y = A_q x + B_q$, 
and the second has the asymptote $y = Ax + B$. 
Let $p_0 = \sqrt{\frac{C_q}{A_q}}$ be the  value of $x$ where $f_q(x)$ has a minimum 
and let $p_1 = \sqrt{\frac{C}{A}}$ be the  value of $x$ where $f(x)$ has a minimum.
If $t_o>0$ and $t_p>0$, then
for each integer $n\geq 1$, the inequality $p_0\leq p_1$ holds.
Moreover, if $n\geq q+1$, then $p_0 <p_1$.
\newtheorem{th1}{Theorem}
\begin{th1}
Let $n\geq 1$ be the number of input elements processed by a bounded 
pipeline. Suppose that the logical delay $t_p$ and the data transformation 
time $t_o$ are both greater than zero. 
For any number $q> 0$, the optimal depth is equal to 
$$
p_{opt} (q, n)=\max\left(\sqrt{ \frac{((n-1) {\rm~mod~}{q}) t_p} 
{ \left(1 + \left[\frac{n-1}{q}\right]\right) t_o} }, \min\left(q,\sqrt{ \frac{(n-1) t_p}{ t_o}}\right)\right).
$$
\end{th1}
{\sc Proof}. 
If $n-1<q$, then $A_q=A$, $B_q=B$, $C_q=C$, and therefore
$p_{opt}(q,n)=p_0=p_1$. In this case the formula is true.

Let $n\geq q+1$.
The function $T_q (x, n)$ takes values
$$T_q (x, n) = 
\begin{cases}
f_q (x), & \text{ for } x\geq q, \\
f (x), & \text{ for } x <q.
\end{cases}
$$
For arbitrary $q> 0$ and $x = q$, the equality $f_q (x) = f (x)$ holds. 
For $x> q$, the inequality $f_q (x)> f (x)$ holds, and for $x <q$, 
the inequality $f_q (x) <f (x)$ is true. Hence, it follows that 
$T_q (x, n) = \max{(f_q (x), f (x))}$ for all $x> 0$ and $q> 0$. 
Wherein the graph of the function $y = f_q (x)$ lies below the graph $y = f (x)$,
 with $0 <x <q$, and the above when $x> q$. 
The point $(q, f_q (q))$ is the unique common point of these graphs. In case $t_o> 0$,
 both functions $f_q (x)$ and $f (x)$ defined on all $x> 0$, have minima.
The inequality $p_0^2 - p_1^2 < 0$ leads to $p_0 <p_1$.
This implies that for every $q> 0$ is one of the cases: (i) $q <p_0$, 
(ii) $p_0\leq q\leq p_1$, or (iii) $p_1 <q$. 
In all cases, for $ x\geq q$ we have $f (x) \leq f_q (x)$, and for $ x\leq q$ we have $f_q (x)\leq f (x)$.
Fig. 2 contains an example showing the graph of the function $T_q (x, n)$, corresponding to the case (i). It is constructed at $q = 5, t_o = 0.001, t_p = 1, n = 50$. In this case, $q <p_0 <p_1$.
 When $x\geq q$, it is true $T_q (x, n) = f_q (x) \geq f_q (p_0) = T_q (p_0, n)$, and if $x\leq q$, then $T_q (x, n) = f (x) \geq f_q ( x) \geq f_q (p_0) = T_q (p_0, n)$.

\begin{center}
\includegraphics[height=50mm]{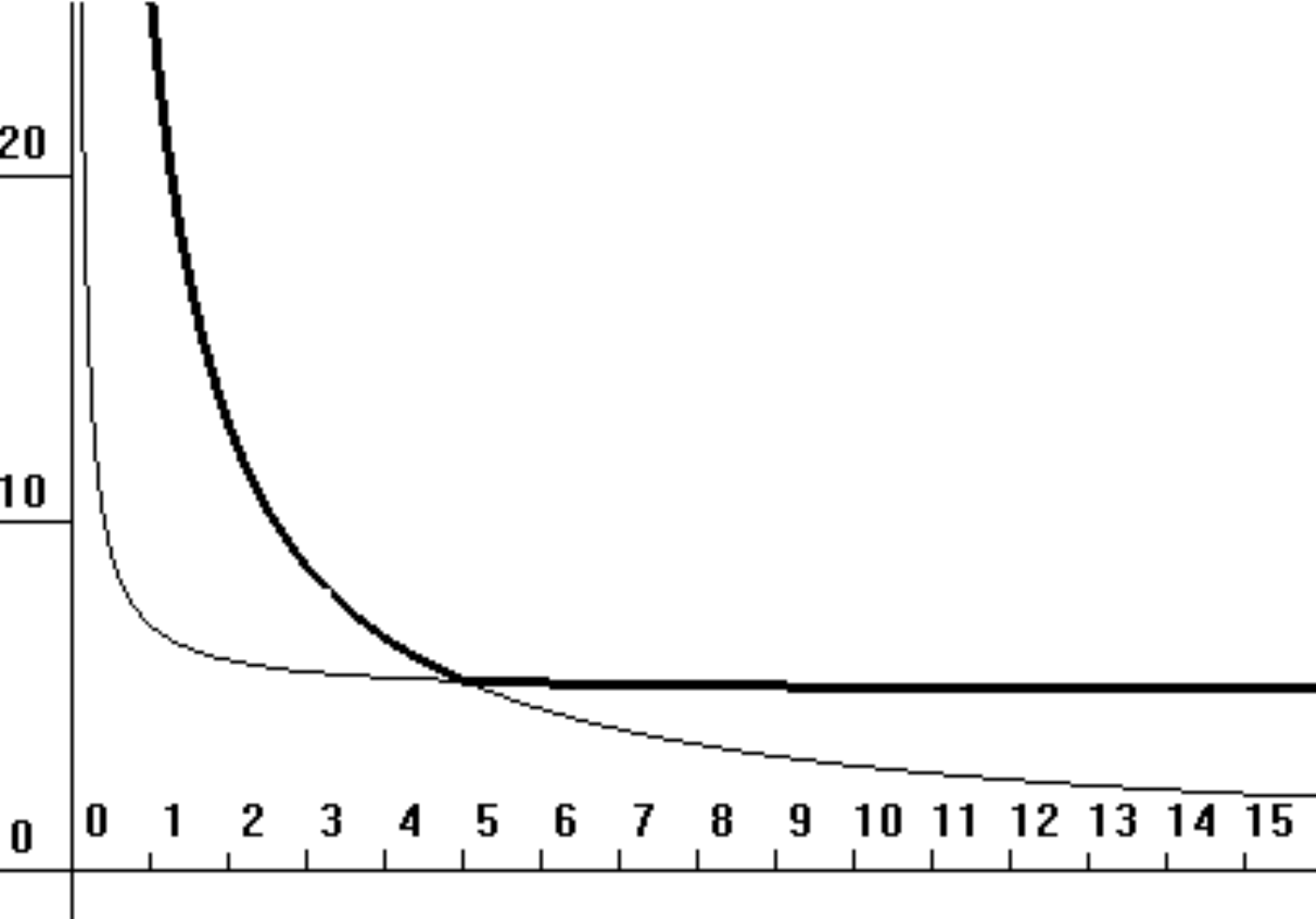}

{\it Fig. 2.}  {The case of $q<p_0$} 
\end{center}
Therefore, in the case (i), the function $T_q(x,n)$ has the minimum in $x=p_0$.

If $p_0\leq q \leq p_1$, then the points of the graph of the function $f_q (x)$, 
as in the first case, will lie above the graph $y = f (x)$ for $x\geq q$
 (see Fig. 1). But in this case $f_q (x)$ increases for $x\geq q$. 
Hence, for $x\leq q$, the values of the function $T_q (x, n)$ are equal to $f (x)$. 
The function $f (x)$ has a minimum for $p_1> q$, so it decreases for $x <q$. 
Consequently, the function $T_q (x, n)$ has a minimum value at $x = q$.

Finally, let $p_1 <q$. The example of this variant is shown in Fig. 3, 
at $q = 12, t_o = 0.5, t_p = 0.5, n = 50$. 
Then, as in the first two cases, $f_q (x) \geq f (x)$ for all $x\geq q$. 
The minima are reached to the left of $q$. 
For $x<q$, the function $f (x)$ has values greater than $f_q(x)$, 
hence, for these $x$, it is true that $T_q (x, n) = f (x)$, 
whence the minimum point of the function $T_q (x, n)$ is coincide with 
the minimum point of the function 
$f (x )$, which reaches a minimum at $x = p_1$.

\begin{center}
\includegraphics[height=50mm]{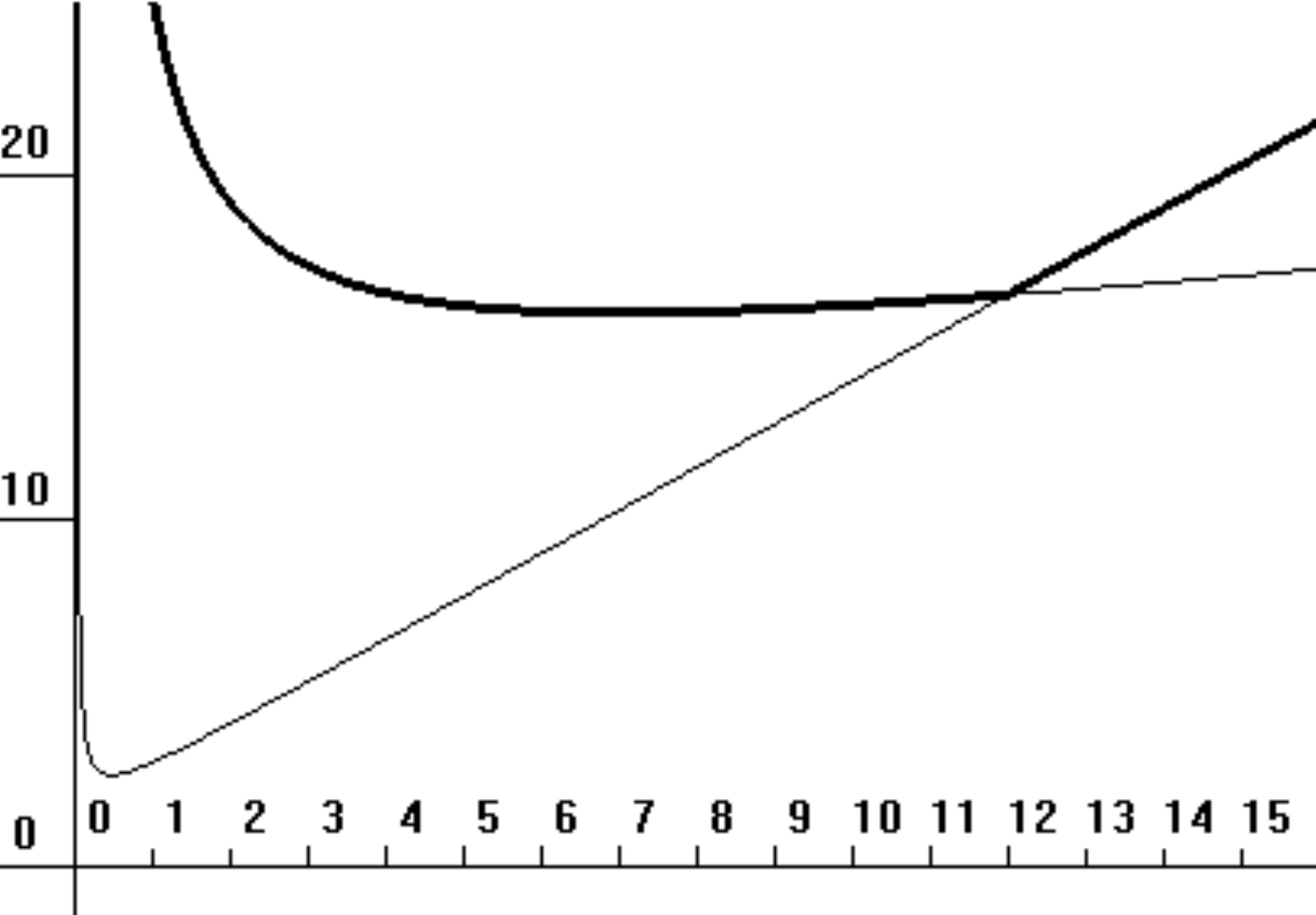}

{\it Fig. 3.} {The case of $p_1<q$} 
\end{center}
The combination of cases (i)-(iii) leads to the desired formula.

\section{Simplified analytical model of a bounded pipeline}

The formula for computing the performance of Proposition 1 has 
the following drawback. 
The optimal depth of the bounded pipeline depends very much on the amount of data, 
even when the data amount differ by $1$. For example, for 
$t_o = 0.02, t_p = 10, q = 15$, in the case of $n = 150$, the optimal depth 
is $p_{opt} (q, n ) = 27$, and if $n = 151$, then $p_{opt} (q, n) = 15$.
In order to correct this defect, we introduce in this section a simplified model.

To this purpose, we first consider a uniform pipeline with hazards consisting of
 $p$ stages. We will assume that each input element is processed continuously 
and at any moment of the time, at least one stage is active. 
The processing time of the first input element is equal to the depth 
of the pipeline. For each input element that is not the first, 
its processing time will be equal to the difference between 
the processing end time of this element and the processing 
end time of the previous element. 
It will be equal to a certain number $j$ of pipeline cycles, 
belonging to the range $1\leq j\leq p$. In particular, if $j=p$, 
then this element causes a restart. In \cite{X2018}, the input element 
is called a hazard of \emph{type} $j$, if the processing time 
is equal to $j$ pipeline cycles, and $j\geq 2$. 
(The pipeline cycle is equal to the delay of the stage and is denoted by $h$.)

Let $b_j$ be the probability that the processing time of the input 
element is $j$. Since at least one stage is active at any time, 
$b_1 +\ldots+ b_p = 1$.

According to \cite[Theorem 1]{X2018}, the processing time of $n$
 input elements by a pipeline is a random value, the mathematical expectation of which 
is equal to
$$
T (p, n) = \left(p + (n-1) (b_1 + 2 b_2 +\ldots + p b_p)\right) h,
$$
where $h$ is the delay time of the stage (pipeline cycle).
A \emph{simplified pipeline} corresponding to a bounded pipeline is 
called a pipeline that allows a single hazard of the type $j = (p-q)^+ +1$ 
with probability $b_j = \frac{1}{q}$. In this case, $b_1 = 1- \frac{1}{q}$. 
The mathematical expectation of the processing time for this pipeline is 
$$
T (p, n) = (p + (n-1) (1-\frac{1}{q} + ((p-q)^+ + 1) \frac{1}{q})) h.
$$

Using the fact that $0\leq x- [x] <1$, we obtain the following assertion.

\newtheorem{prop2}[prop1]{Proposition}
\begin{prop2}
For $p\leq q$, the equality $T (p, n) = T_q (p, n)$ holds, and for 
$p> q$, the following inequalities hold:
$$
0\leq T (p, n) -T_q (p, n) <(p-q)^+ (t_o + \frac{t_p}{p}).
$$
\end{prop2}

\newtheorem{cor1}{Corollary}
\begin{cor1}
The optimal depth of a simplified pipeline, corresponding to a 
bounded pipeline, equals
$$
\widetilde{p}_{opt} (q, n) = \min\left(q, \sqrt{\frac{(n-1) t_p}{t_o}}\right).
$$
\end{cor1}
{\sc Proof}. Graph of the function
$$
T (x, n) = 
\begin{cases}
(1 + \frac{n-1}{q}) x (t_o + \frac{t_p}{x}), & \text{for $x\geq q$,} \\
(x + n-1) (t_o + \frac{t_p}{ x}) , & \text{for $x<q$.}
\end{cases}
$$
consists of a part of the hyperbola and part of the ray emerging from the 
point $(0, (1+ \frac{n-1}{q}) t_p)$. If the ray intersects the hyperbola 
to the left side of the minimum point, 
then $p_{opt} (q, n) = q$. If on the right, then the function $y = T (x, n)$ gets 
the minimum value at the point 
corresponding to the minimum value of the hyperbola.

\section{Optimal depth of the bounded pipeline with restarts}

Our next problem is to find a formula for optimal depth 
of a bounded pipeline that accepts random restarts with a given probability.
This depth should not change too much when the data amount changes are 
small. To solve this problem, we again apply a simplified pipeline
corresponding to a bounded one.

Consider a pipeline of depth $p$ that allows two hazards.
Its first hazard is restart with probability $b_p=b$. The second has type 
$(p-q)^+ +1$. It can not occur together with a restart, whence its probability 
is equal to $(1-b)\frac{1}{q}$. The probability of processing a data element 
in one pipeline cycle is $b_1= (1-b)(1-\frac{1}{q})$. 
An analytical model for the processing time of $n$ input elements using
a simplified pipeline will be described by  the formula
$$
T(p,n,b)= \left(p+(n-1)\left((1-b)(1-\frac{1}{q})+((p-q)^+ +1)(1-b)\frac{1}{q}
+bp\right)\right) h.
$$

\newtheorem{th2}[th1]{Theorem}
\begin{th2}
Optimal depth of simplified pipeline with restarts equals
$$
\widetilde{p}_{opt}(q,n,b) = \min\left(q, \sqrt{\frac{(1-b) t_p}{(\frac{1}{n-1}+ b)t_o}}\right).
$$
\end{th2}
{\sc Proof.} The function $T(x,n,b)$ has the values
$$
T(x,n,b)=
\begin{cases}
\left(1+(n-1)(\frac{1-b}{q}+b)\right)x(t_o+\frac{t_p}{x}), & \text{for $x\geq q$},\\
(x+(n-1)(1-b+bx))(t_o+\frac{t_p}{x}), & \text{for $x\leq q$}.
\end{cases}
$$
Its graph consists of a part of the hyperbola lying in the first quarter 
and a part of the ray emerging from the point
$\left(0, \left(1+(n-1)(\frac{1-b}{q}+b)\right)t_p\right)$.
The ray intersect the hyperbola at $x=q$.
The abscissa of the lower point of the hyperbola equals 
$p_1= \sqrt{\frac{(1-b) t_p}{(\frac{1}{n-1}+ b)t_o}}$.
If $q\leq p_1$, then the function $T(x,n,b)$ decreases on the 
interval $(0,q)$ and increases for $x>q$, and hence $T(x,n,b)$ has 
the minimum at $x=q$. If $q\geq p_1$, then for $x\geq q$ it is increasing
and we obtain that $T(x,n,b)$ has minimum at $x=p_1$.

The formula obtained generalizes to bounded pipelines the formula from
\cite{X20182} which
refines the Dubey and Flynn formula from \cite{dub1990}.

\section{Computer modeling of bounded pipeline}

We use two methods of simulating the operation of bounded pipelines. 
Both methods are suitable for measuring performance.
The first method is based on the use of multithreaded pipelines operating 
under the control the operating systems Windows. 
Each stage of a multithreaded pipeline is implemented as a thread that contains 
a loop consisting of reading data from the input channel, performing a stage operation, 
and writing the results to the output channel.
The operation is simulated by waiting time operator of the delay time of the stage
 or the time of recording in the lock. A channel is defined as an object 
of a class consisting of a queue and operations for writing and reading queue elements. 
Its software implementation is described in the preprint \cite{X2004}. 
Fig. 4 shows a graph of the processing time of $n=20$ elements, 
obtained with a multithreaded pipeline, with the number of processors 
$q = 5$, the logical delay $t_p = 100$ milliseconds, and the write time 
in the lock $t_o = 3$. Small circles indicate the values obtained experimentally. 
The graph obtained by formula (1) is pictured by lines.
Experimental value of the optimal depth equals $6$.
Theoretical $p_{q,n}\approx 5.8$. Moreover, the integer number for which 
$T_q(x,n)$ is minimal equals $6$. 
  
\begin{center}
\includegraphics[height=50mm]{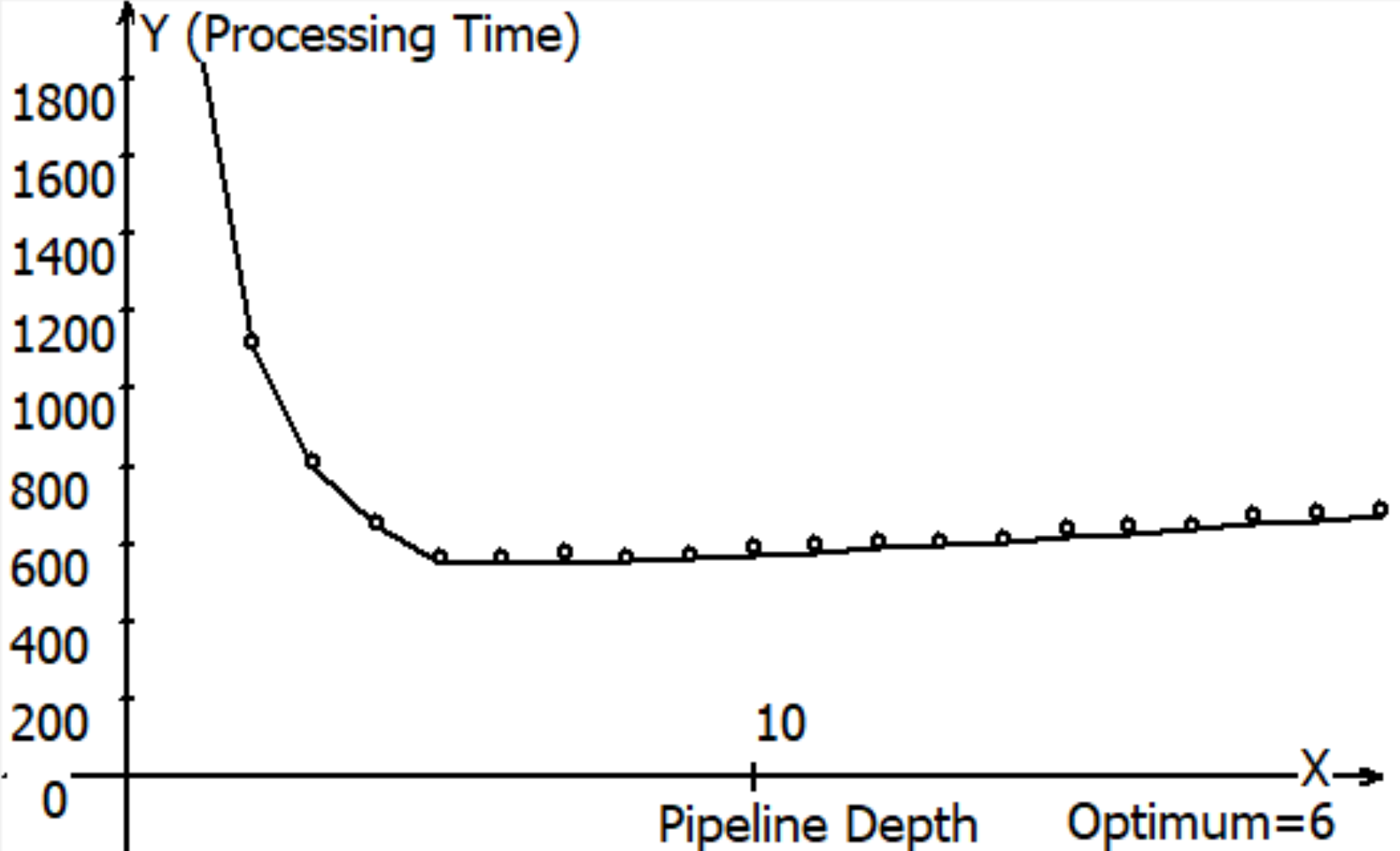}

{\it Fig. 4.} {The result of a multithreaded pipeline}
\end{center}

The second method is based on the theory of trace monoids and is described by Diekert 
\cite{die1990}. 
An arbitrary program is decomposed into a sequence of operations having an execution 
time equal to one clock cycle. If the operations can be performed in parallel, 
then they are treated as permutation. By rearranging the operations,
 we get the maximum block that can be executed during the first measure. 
We execute this block and proceed to the operations that remained. 
Using permutations of independent operations, we again select the maximum block 
that will be executed during the second measure, etc. 
These blocks constitute the so-called normal form of the Foata, 
and their number is the height of the normal form. 
In particular, each pipeline can be associated with a sequence of operations, 
and get its normal form. In \cite{X2016} 
this method was applied to a bounded pipeline. 
In this case, the blocks of normal form should not exceed the number of active stages. 
Fig. 5 shows the result of an experiment based on this method for the case when 
the data amount is $n = 50$, the number of stages is $p = 10$, the number of active stages is $q = 5$.

\begin{center}
\includegraphics[height=50mm]{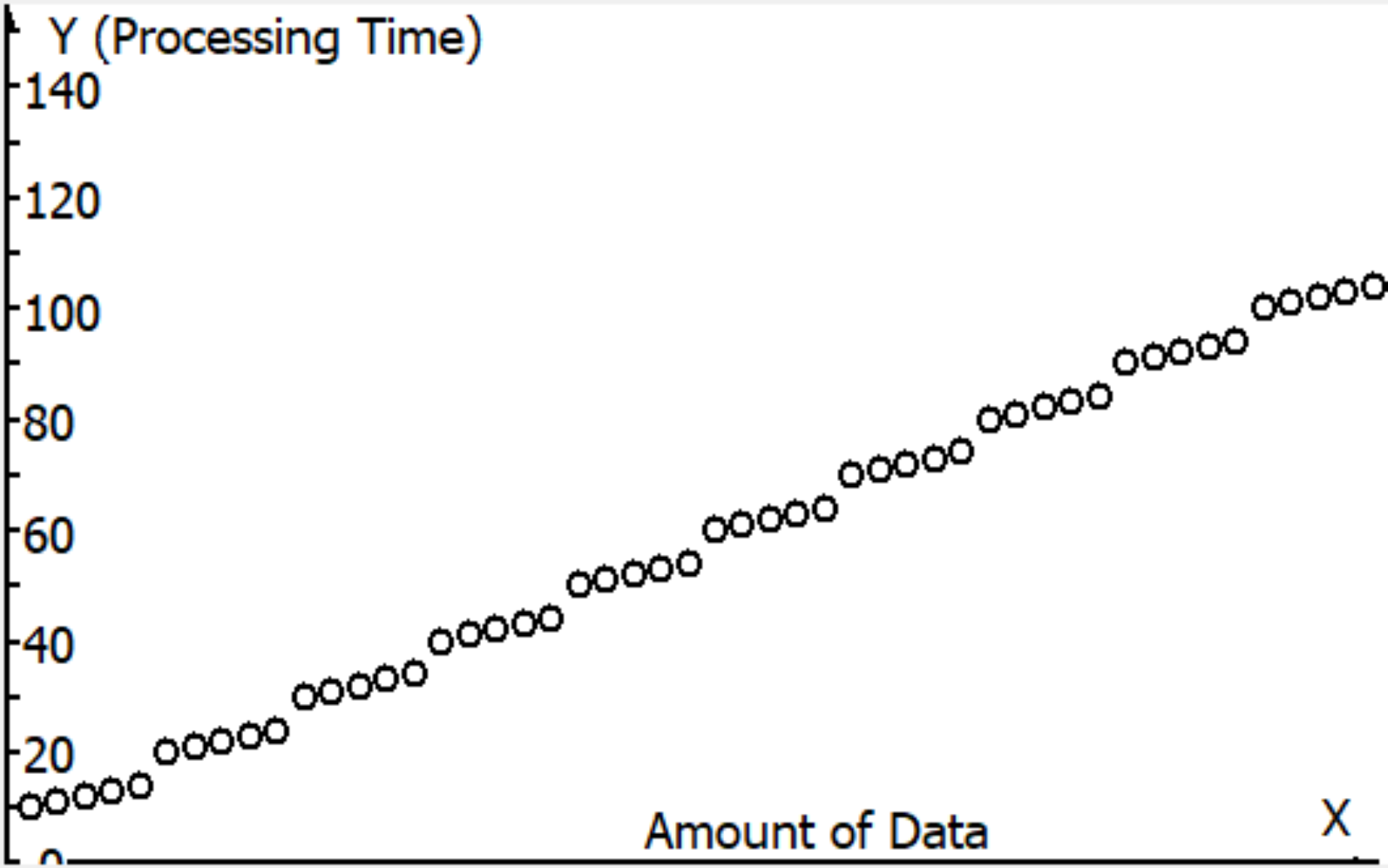}

{\it Fig. 5.} {Time processing in pipeline cycles}
\end{center}

\section{Conclusion}
The pipeline processes the finite sequences of data elements. Theorem 1 shows that in the case 
when these arrays have a constant length $n$, it is reasonable to take this 
length into account. If the data volumes differ, then it is better 
to use Corollary 1. But for this it is necessary to know the density 
of the input data stream, determined by the probability of restart.
In the future, the extension of Corollary 1 and the results of \cite{X20182}, 
the optimal depth that minimizes the processing time of a given amount, to bounded pipelines, and the study of bounded pipelines with other hazards. In addition, it is possible to generalize Proposition 1 to bounded pipelines, 
the stage delays of which are not equal to each other. This should result in the calculation 
of the minimum number of functional devices and other useful properties 
of uneven bounded pipelines.


\section*{Acknowledgment}

This work was performed as a part of the Strategic Development Program at the National
Educational Institutions of the Higher Education, N 2011-PR-054.

\end{document}